\documentstyle[preprint,aps,prb]{revtex}

\begin{document}
\title{Single-electron-parametron-based logic devices}
\author{Alexander N. Korotkov$^{1,2}$ and Konstantin K. Likharev$^{1}$}
\address{$^1$Department of Physics, State University of New York, 
Stony Brook, NY 11794-3800 \\
and 
$^2$Nuclear Physics Institute, Moscow State University, Moscow 119899, 
Russia.}
\date{\today }
\maketitle

\begin{abstract}
We analyze the operation of the wireless single-electron logic family based
on single-electron-parametron cells. Parameter margins, energy dissipation,
and the error probability are calculated using the orthodox theory of
single-electron tunneling. Circuits of this family enable quasi-reversible
computation with energy dissipation per bit much lower than the thermal
energy, and hence may circumvent one of the main obstacles faced by
ultradense three-dimensional integrated digital circuits.
\end{abstract}


%\pacs{73.23.Hk; 73.40.Gk}

%\preprint{Submitted to J. Appl. Phys.}

\narrowtext

%\vspace{1ex}

\section{Introduction}

Single-electron tunneling \cite{Av-Likh,SCT} in systems of small-capacitance
tunnel junctions has attracted much attention during the last decade because
of both the interesting physics and the possibility of useful devices. Today
the basic physics of single-electron tunneling is sufficiently well
understood, and its possible applications (for reviews see, e.g., Refs.\ 
\cite{Av-Likh-log} and \cite{Kor-rev}) are becoming an important issue. The
practical value of several analog applications has already been proven: for
example, the single-electron transistor \cite{Ave-Lik-85,Fulton-D} as a
highly sensitive electrometer,\cite{Likh-87} the single-electron pump \cite
{Pothier-pump} as a standard of dc current, \cite{Keller} and the array of
tunnel junctions \cite{Bakh-89} as an absolute thermometer.\cite{Pekola}

In the field of digital electronics, however, only rudimentary devices have
been demonstrated so far. \cite
{LaFarge-trap,Likh-trap,Yano-mem,Nakazato-mem,Min-mem,Fuji-mem,NEC-mem} The
recent invention \cite{Q0indep} of hybrid circuits based on
single-electron-transistor readout from floating-gate memory cells
presumably opens a way to room-temperature {\it memories} with density
beyond $10^{11}$ bit/cm$^{2}$. (This concept has been supported by recent
experiments \cite{NEC-mem} with the first low-temperature prototypes of such
memory cells). The potential of single-electronics in {\it logic} circuits,
however, remains uncertain due to substantial problems which have to be
solved for both types of single-electron logic devices suggested so far.

Devices of the first type use single-electron transistors instead of field
effect transistors in ``voltage-state'' circuits similar to conventional
digital electronics. \cite{Likh-87} In these devices the effects of charge
quantization are restricted to the interior of the single-electron
transistors. As a result, the design of such circuits (including
complementary \cite{Tucker}) is relatively straightforward. They suffer,
however, from a relatively high static power consumption. \cite
{Kor-trans,Chen}

The second possible way (which allows to reduce the power dissipation), is
to code bits directly by single electrons everywhere in the circuit. \cite
{Likh-Sem,Likh-Polon,Naz-Vysh,Nakazato-log,Kor-isl,Param1,Param2,Ancona,%
GSC,Lent-Tou}
(A different, interesting approach based on rf parametric excitation of
single-electron-tunneling oscillations has been suggested by Kiel and
Oshima; \cite{Kiehl} unfortunately their original proposal seems to have run
into considerable implementation problems \cite{Oshima}). It is important
that these ``charge-state'' logic circuits do not necessarily require long
wires and can be implemented using only small conducting islands separated
by tunnel barriers, with the necessary power and timing provided by an
external rf electric field. \cite{Kor-isl} Recently we suggested \cite
{Param1} a new device for charge-state, wireless logic circuits - the Single
Electron Tunneling Parametron (or SET Parametron). In comparison with the
wireless single-electron devices suggested earlier,\cite{Kor-isl} the SET
Parametron may have wider parameter margins and extremely low power
dissipation, lower than the thermodynamic ``limit'' of $k_{B}T\ln 2$ per
bit. \cite{Param2}

In this paper we present the results of a detailed analysis of the SET
Parametron and basic logic circuits using this device. The major
characteristics of the SET Parametron, including its parameter margins,
speed, error rate, and power consumption, will be discussed.

\section{SET Parametron: The Basic Idea}

The basic cell of the new logic family, the SET Parametron, consists of
three conducting islands - see Fig.\ \ref{three-isl}a. (In our numerical
calculations we have assumed the islands to be of spherical shape, with $R$
being the radius of the edge islands 1 and 3, and $r$ the radius of the
middle island 2; this assumption is, however, not important for the device
operation.) The middle island is slightly shifted off the line passing
through the centers of the edge islands. We will identify this line with 
$x$-axis, and the direction of the middle island shift with $y$-axis (which 
is the symmetry axis of the cell). Electrons can tunnel through small gaps
between the middle and edge islands, but not directly between the edge
islands because of their much larger spacing (Fig.\ \ref{three-isl}a).

Let us assume that the cell as a whole is charged by one uncompensated
electron. (This assumption makes the explanation of the operation principle
simpler; later we will consider the more natural case of an initially
neutral cell.) If the cell is biased with a sufficiently strong ``clock''
electric field $E_{c}>0$ directed along axis $y$, the electron is obviously
located at the central island.\cite{sign} Now let the field be decreased
gradually so that eventually it changes direction to negative. At some
moment the electron will have to tunnel to one of the edge islands because
these states become energetically preferable. Because of the geometrical
symmetry, the choice of island (left or right) will be random, i.e. the
charge symmetry of the system will be broken spontaneously.

However, if there is a weak ``signal'' field $E_{s}$ along direction $x$
(this field may be applied, for example, by the neighboring similar cell),
the final position of the electron will depend on the sign of $E_{s}$. A
natural way to discuss this effect is to say that the signal field $E_{s}$
creates an energy difference $\Delta $ between the electron states in
islands 1 and 3, and that the electron prefers to tunnel into the island
with the lowest energy state (Fig.\ \ref{three-isl}b). If now the clock
field $E_{c}$ becomes negative and large, it creates a high energy barrier 
$W$ (see Fig.\ \ref{three-isl}b) between the edge islands, so that the
electron is reliably trapped in the island it has tunneled to, regardless of
further changes of the signal field $E_{s}$.

Thus if fluctuations in the system are low enough, and the clock field
changes slowly enough, even a small field $E_{s}$ of the proper sign at the
decision-making moment (when $W(t)$=0) is sufficient to ensure a certain
robust final polarization of the cell. This process can be interpreted as a
reliable recording of one bit of information; for example, the electron on
the right island may mean logical ``1'' while the electron on the left
island encode logical ``0''. Now the dipole moment of the cell can in turn
be used to produce the signal field to control the other cells during their
decision-making moments, and hence determine their information contents (see
Section V below).

This ``parametric'' amplification of signal, which gave the SET Parametron
its name, is quite similar to the operation of a Josephson junction device
called Parametric Quantron \cite{Likh-PQ,Likh-PQ2} (re-invented later as the
``Flux Quantum Parametron'' \cite{Goto-PQ}), and in a broader sense to
rf-driven parametrons. \cite{Param-book,Kiehl} The main difference is that
all the previously analyzed parametrons are described by a continuous degree
of freedom (e.g., the Josephson phase $\phi $ in the Parametric Quantron).
This variable can take any value, and the discrete information states
correspond to energy minima on the $\phi $ axis. On the other hand, in the
SET Parametron with low tunnel conductances, the possible states are
discrete (the electron has to be definitely on one of the islands at each
particular instant). \cite{higher-states}

\section{SET Parametron: Quantitative Analysis}

Let us consider the operation of one SET Parametron cell using the orthodox
theory \cite{Av-Likh} which is adequate for the description of
single-electron tunneling in systems with many background electrons in each
island and sufficiently low conductance $G$ of tunnel junctions ($G\ll
e^{2}/\hbar $). In this theory the electron is always localized in one of
the islands, and the rate $\Gamma $ of each tunneling event is solely
determined by the corresponding decrease $W=-\Delta \varepsilon $ of
electrostatic energy $\varepsilon $ of the system. Using elementary
electrostatics, we obtain the following expressions for the energy decreases
corresponding to four possible tunneling events in the SET Parametron
(tunneling to/from each of the edge islands is denoted with signs $+$ and 
$-$, respectively):

\begin{eqnarray}  
W^\pm_l &=& \pm e \, (\phi_2-\phi_1) - W_{12}, \,\,\,
\nonumber \\ 
 W^\pm_r &=& \pm e \,
(\phi_2-\phi_3) - W_{23} \, ,
\label{gain} \end{eqnarray}

\begin{equation}
\phi_i = \sum_{j=1}^3 \, (C^{-1})_{ij} q_j - {\vec E} {\vec D}_{i} \, ,
\end{equation}
\begin{equation}  \label{Wij}
W_{ij} = \frac{e^2}{2} \left[ (C^{-1})_{ii} + (C^{-1})_{jj} -2 (C^{-1})_{ij} 
\right] \, ,
\end{equation}
\begin{equation}
q_i=q_{i0}+n_i e \, .
\end{equation}

Here $l$($r$) means left (right) junction, $\phi _{i}$ is the electrostatic
potential of $i$-th island (for numbering, see Fig.\ \ref{three-isl}a), 
$W_{12}$ and $W_{23}$ are the Coulomb blockade energies, and $C^{-1}$ 
is the inverse capacitance matrix. Vectors 
$e{\vec{D}}_{i}=(eD_{ix},\,eD_{iy},\, eD_{iz})$ are the dipole moments 
of islands when charged with single
electrons, ${\vec{E}}$ is the external electric field (accepted to be
uniform) and $q_{i}$ is the total electric charge of $i$-th island, while 
$q_{i0}$ is the uncompensated part of its background charge.

In the limit when the island radii are smaller than the spacing between
islands, the following simple approximation for the capacitance matrix
elements can be used: 
\begin{eqnarray}
&&(C^{-1})_{11}=(C^{-1})_{33}=\frac{1}{4\pi \epsilon \epsilon _{0}}\,
\frac{1}{R}\,;\,\,\,\,(C^{-1})_{22}=\frac{1}{4\pi \epsilon 
\epsilon _{0}}\,\frac{1}{r}\,;  \nonumber \\
&&(C^{-1})_{ij}=\frac{1}{4\pi \epsilon \epsilon _{0}}\,\frac{1}{s_{ij}} 
\,,\,\,\,i\neq j\,;  \nonumber \\
&&s_{ij}=\left[ (x_{i}-x_{j})^{2}+(y_{i}-y_{j})^{2}+(z_{i}-z_{j})^{2}\right]
^{1/2};\,\,  \nonumber \\
&&{\vec{D}}_{i}={\vec{r}}_{i},  \label{approx}
\end{eqnarray}
where ${\vec{r}}_{i}=(x_{i},y_{i},z_{i})$ is the $i$-th island center, and 
$\epsilon $ is the dielectric constant of the dielectric environment. In the
case of comparable radii and island spacing, $C_{ij}$ have to be calculated
numerically, for example, using the method of multiple electrostatic images
(see Appendix A). The results show that approximation (\ref{approx}) gives
accuracy better than 10\% if the spacing between the islands is larger than
the largest island's radius.

In our initial analysis (until Section VII) we will assume that temperature 
$T$ is low and that the electric fields are changing slowly (adiabatically).
In this case the full result of the orthodox theory, 
\begin{equation}
\Gamma =\frac{G\,W}{e^{2}\,[1-\exp (-W/k_{B}T)]}  \label{Gamma}
\end{equation}
is reduced to the simple rule that the system always follows the local
minimum of the total electrostatic energy.

Figure \ref{diag}a shows the phase diagram of stationary charge states of a
cell charged by one uncompensated electron. Each state is locally stable
within a diamond-shape region corresponding to two conditions: 
\begin{equation}
|e(\phi _{2}-\phi _{1})|<W_{12}\,,\,\,\,\,|e(\phi _{2}-\phi _{3})|<W_{23}\,,
\label{diamond}
\end{equation}
being valid simultaneously, so that no tunneling event is possible, 
$W_{l,r}^{\pm }<0$. Within the small-island approximation (\ref{approx}) 
the position of the diamond center is 
\begin{eqnarray}
E_{s} &=&\frac{1}{4\pi \epsilon \epsilon _{0}}\,\frac{1}{2}\,\frac{1}{d}
\,(n_{1}-n_{3})\,e\left( \frac{1}{2d}-\frac{1}{R}\right) ,  \nonumber \\
E_{c} &=&\frac{1}{4\pi \epsilon \epsilon _{0}}\,\frac{1}{2}\,\frac{1}{b} 
\left[ (n_{1}+n_{3})\,e\left( \frac{2}{s}-\frac{1}{2d}-\frac{1}{R}\right)
\right. \nonumber \\
&& \left. +2n_{2}e\left( \frac{1}{r}-\frac{1}{s}\right) \right] ,  
\label{center} \end{eqnarray}
while the diamond width and height are 
\begin{eqnarray}
\delta E_{s} &=&\frac{2}{e}\,\frac{1}{d}\,W_{12}\,,\,\,\,\,\,\delta E_{c}= 
\frac{2}{e}\,\frac{1}{b}\,W_{12}\,,  \nonumber \\
W_{12} &=&W_{23}=\frac{e^{2}}{2}\,\frac{1}{4\pi \epsilon \epsilon _{0}}
\,\left( \frac{1}{r}+\frac{1}{R}-\frac{2}{s}\right) \,  \label{size}
\end{eqnarray}
(here $s\equiv s_{12}=s_{23}=(d^{2}+b^{2})^{1/2}$ is the distance between
island centers -- see Fig.\ \ref{three-isl}a).

The diamonds are periodic along both axes $E_{s}$ and $E_{c}$, with periods 
\begin{eqnarray}
\Delta E_{s} &=&\frac{1}{4\pi \epsilon \epsilon _{0}}\,\frac{1}{d}\,e\left( 
\frac{1}{R}-\frac{1}{2d}\right) ,  \nonumber \\
\Delta E_{c} &=&\frac{1}{4\pi \epsilon \epsilon _{0}}\,\frac{1}{b}\,e\left( 
\frac{1}{R}+\frac{2}{r}+\frac{1}{2d}-\frac{4}{s}\right) .  \label{period}
\end{eqnarray}
Equations (\ref{center})--(\ref{period}) are valid even if the approximation
(\ref{approx}) is not used; however, in this case $1/R$, $1/r$, $1/s$, $1/2d 
$ (divided by $4\pi \epsilon \epsilon _{0}$) need to be replaced by 
$C_{11}^{-1}$, $C_{22}^{-1}$, $C_{12}^{-1}$, $C_{13}^{-1}$, respectively, 
and also $b$ and $d$ by $D_{2y}-D_{1y}$ and $D_{2x}-D_{1x}$ (the 
corresponding
equations are direct consequences of Eqs.\ (\ref{gain})--(\ref{Wij}) and 
(\ref{diamond})).

The period along axis $E_{s}$ corresponds to the transformation 
$(n_{1},n_{2},n_{3})\rightarrow (n_{1}-1,n_{2},n_{3}+1)$, while the period
along $E_{c}$ axis corresponds to $(n_{1},n_{2},n_{3})\rightarrow
(n_{1}-1,n_{2}+2,n_{3}-1)$. Notice that these transformations allow only
even-number changes at the middle island; the ``odd'' set of diamonds can be
obtained from the ``even'' set by the shift 
$(\Delta E_{s}/2,\Delta E_{c}/2)$.

Figure \ref{diag}a shows the results of more exact, numerical calculations
of the phase diagram for a particular cell; however, they are quite close to
the approximate analytical results (\ref{center})--(\ref{period}). (The
approximation accuracy is better than 10\% and it becomes even better than
1\% after the normalization by $\Delta E_{s}$ and $\Delta E_{c}$.) For
example, the minor offset of the diamond corners from the horizontal axis,
which arises due to deviations from the analytical approximation, is hardly
visible.

The most important qualitative feature of the phase diagram is the existence
of bistability regions (shaded in Fig.\ \ref{diag}a) where either of two
charge states is locally stable. We will refer to the SET Parametron in
these regions as being in an ON state, while the remaining part of the phase
diagram corresponds to one of the possible OFF states of the system. If
signal field $E_{s}$ is low (as we suppose in the following), only the set
of diamonds along axis $E_{c}$ may be implemented. The arrow in Fig.\ \ref
{diag}a shows a possible way (described qualitatively in Sec.\ II above) of
switching the cell from a monostable OFF state to a bistable ON state by
changing the clock field $E_{c}$. It is clear that the sign of a small 
$E_{s} $ field determines which diamond boundary will be crossed first, and
hence determines the charge state of the system within this bistable region.

Notice that crossing any phase boundary outside of the bistable regions
involves one tunneling event with negligible energy difference $W$. In the
adiabatic limit this transition is reversible, and the associated energy
loss is infinitesimal (see Sec.\ VII). However, leaving the bistable state
from the state with higher energy results in two sequential tunneling events
and leads to a finite energy loss. This irreversible switching should be
avoided in reversible computation. Below we will show that SET Parametron
circuits can operate using exclusively reversible transitions.

\section{Single Exciton Parametron}

While the case of the cell charged by one electron is the simplest for
understanding the device operation, a more natural option is the initially
electrically neutral cell: $n_{1}+n_{2}+n_{3}=0$. (This variety of SET
Parametrons may be called the ``Single Exciton Parametron'', since digital
bits in it are presented by electrostatically bound electron-hole pairs,
although the physics of this bound state is rather different from that of
the usual excitons in semiconductors.) Figure \ref{diag}b shows the phase
diagram of such a cell with the same geometrical parameters as in Fig.\ \ref
{diag}a. The only difference between the diagrams is a change in the charge
states labeling and a fixed shift along $E_{c}$ axis, so that the diagram in
Fig.\ \ref{diag}b is symmetric about both axes. (Actually, {\it any} change
of the initial charge of the cell is equivalent to a fixed offset of the
fields $E_{c}$ and $E_{s}$).

Due to its field sign symmetry, the Single Exciton Parametron allows two
natural modes of operation. The first is illustrated by the arrowed
rectangle in the center of the diagram (instead of the thick line in Fig.\ 
\ref{diag}a we have drawn the rectangle in Fig.\ \ref{diag}b to emphasize
that $E_{s}$ may shift during the operation). The periodic clock field 
$E_{c}(t)$ causes two switchings OFF$\rightarrow $ON and two switchings 
ON$\rightarrow $OFF per period. Another possible mode of operation is 
shown by
the smaller rectangle centered at $E_{s}$=$\Delta E_{s}/2$. It also provides
two pairs of ON/OFF switchings per period of the clock field $E_{c}(t)$,
with the bistable state implemented at small fields. In our analysis we will
concentrate on the former operation mode because it does not require the
additional dc bias of the signal field, though the latter mode may allow
wider parameter margins.

Figures \ref{diag}a and \ref{diag}b show the phase diagrams of SET
Parametrons for a particular set of geometric parameters. Changing the
parameters may result not only in quantitative but also qualitative changes
of the diagram. In particular, regions with three stable charge states
appear if the ratio $\delta E_{s}/\Delta E_{s}$ of the diamond width and
period becomes larger than 2; if the ratio is larger than 3, four states can
coexist, and so on. Geometrically the multi-stability requires a smaller
middle island to increase the Coulomb barrier for the electron transfer
between the edge islands via the middle island. For the sake of simplicity
we will limit ourselves to the bistable case. \cite{inversion}

\section{Shift register}

Figure \ref{SR-idea} shows the standard scheme of operation of a shift
register using parametron-type binary cells.\cite{Param-book} In this
figure, each symbol represents a parametron cell at some particular phase of
the periodic switching process. Dash means the monostable (OFF) state, while 
{\bf I}$_{{\bf i}}$ means the $i$-th data bit which can be either ``0'' or
``1''. Evolution of the symbols shows the bit flow when the cells are
periodically switched ON and OFF using periodic clock signals which are
shifted by $2\pi /3$ for each next stage of the register.

In the initial state (upper line) each bit is encoded by the corresponding
ON state of only one parametron. The signal field created by this cell is
exerted on both neighboring cells, initially in their OFF state. Due to this
field, when the right neighboring cell is switched from OFF to ON by the
clock field (second line in Fig.\ \ref{SR-idea}), its logic state will be
determined by that of the initial cell. Now when the information has been
copied to the right cell, the initial cell can be switched OFF (third line
of Fig.\ \ref{SR-idea}). Thus, the information has been moved by one cell
(one register stage), and the process may be repeated again (lines 4 to 7)
and again. Notice that unidirectionality of the bit propagation is achieved
by the ``running wave'' of the clock while the structure itself is
1D-isotropic.

During each period of the three-phase clock, the information is shifted by
three stages, so that we need 3 cells per bit of information. It is
straightforward to increase the number of phases beyond 3. Typically this
would make the system more robust, but require proportionally more hardware
per bit. (More generally, the number of cells per bit has not to be integer:
by changing the clock phase shift it can have any value larger than 2).

Figure \ref{SET-sr} shows a natural implementation of shift register using
Single Exciton Parametron cells.\cite{Param1} Each next cell, the direction
of the middle island shift is turned by $\pi /3$ (within plane $y-z$). The
system is driven by the clock field vector ${\vec{E}(t)}$ rotating in the
same plane. This rotation provides the shift of $E_{c}$ (which is now the
component of the clock field in the plane of the corresponding cell) by 1/6
of the clock period for each next cell. When a cell is in ON state, its
dipole electric moment creates the signal field $E_{s}$ which is especially
strong for its nearest neighbors, and thus determines the direction of
electron tunneling when one of the neighbors in turned ON.

The operation of the system is illustrated in Fig.\ \ref{sr-diag}. Three
sine lines show the in-plane components of the clock field for each of three
neighboring cells. Since the operation cycle corresponds to the central
rectangle shown in Fig.\ \ref{diag}b, it differs slightly from the
traditional mode shown in Fig.\ \ref{SR-idea}: each cell is switched OFF and
ON twice per clock period.\cite{doubling} In the simplest case of weak
coupling when the signal field $E_{s}$ is much smaller than the clock field
amplitude, switchings happen when the clock field $E_{c}$ crosses one of the
thresholds $\pm \delta E_{c}/2$ and the cell enters or leaves the bistable
region (shaded in Fig.\ \ref{sr-diag}). For correct operation of the shift
register, switching OFF$\rightarrow $ON of a cell (see, e.g., point $S$ at
the thicker middle curve) should occur when the previous neighboring cell is
in ON state while the next neighboring cell is in OFF state, just as shown
in Fig. \ref{sr-diag}.

In order to find the width of parameter window corresponding to correct
operation of the parametron, let us first neglect the inter-cell
interaction. Using the operation diagram shown in Fig.\ \ref{sr-diag} and
the phase diagram of Fig.\ \ref{diag}b it is straightforward to obtain that
the amplitude $A$ of the rotating clock field has to satisfy the following
three conditions: 
\begin{eqnarray}
&&A\sin (\frac{\pi }{2}-\frac{\theta }{2})>\frac{\delta E_{c}}{2}\,, 
\nonumber \\
&&A\sin (\frac{\theta }{2})<\frac{\delta E_{c}}{2}\,,  \nonumber \\
&&A<\Delta E_{c}-\frac{\delta E_{c}}{2}\,.  \label{amplitude}
\end{eqnarray}
where $\theta $ is the clock phase shift between the adjacent cells. The
first condition is necessary to have the signal source cell still in ON
state when a recipient cell switches from OFF to ON. The second relation
ensures that at that instant the next cell is already in OFF state. Finally,
the third inequality means that the clock field does not reach the next
diamond at the $E_{c}$ axis - see Fig.\ \ref{diag}b and Eqs.\ (\ref{size})
and (\ref{period}), so that no undesirable charge states appear during the
operation.

Equations (\ref{amplitude}) can be reduced to the following equation for the
one-side margin $\xi $ of the field amplitude, $\xi \equiv
(A_{max}-A_{min})/(A_{max}+A_{min})$: 
\begin{equation}
\xi =\min \left[ \tan (\frac{\pi }{4}-\frac{\theta }{2}),\,\,\frac{2\Delta
E_{c}/\delta E_{c}-1-1/\cos (\theta /e)}{2\Delta E_{c}/\delta E_{c}-1+1/\cos
(\theta /e)}\right] .  \label{margin}
\end{equation}
Within the approximation (\ref{approx}) the ratio $\Delta E_{c}/\delta E_{c}$
is equal to $2-(1/R-1/2d)/(1/R+1/r-2/s)$; this ratio increases when the
radius of the middle island decreases. In the general case when the terms 
$1/R$, $1/r$, $1/2d$, and $1/s$ are replaced for the elements of the inverse
capacitance matrix, the upper bound for $\Delta E_{c}/\delta E_{c}$ is equal
to 2, while the lower bound is equal to 1.\cite{inversion}

For the particular case shown in Fig.\ \ref{SET-sr} we have 
$\theta =2\pi /6$, hence, Eq.\ (\ref{margin}) gives the maximum margin 
of $\xi =\tan (\pi /12)\simeq 27\%$ for the amplitude $A$ of 
the rotating clock field if $\Delta E_{c}/\delta E_{c}>1.5$, 
so that the minimum in Eq.\ (\ref{margin})
is determined by the first term. (In order to satisfy this additional
condition the middle island should be sufficiently smaller than the edge
islands.)

The above single-cell analysis is only an approximation, because in the real
structure all the cells interact with each other and also the presence of
neighboring cells somewhat changes the cell electrostatics. This is why we
have carried out extensive numerical simulations of reasonably long shift
registers (18 three-island cells). For a given geometry of the circuit we
first calculated numerically the inverse capacitance matrix $C_{ij}^{-1}$
and the vectors ${\vec D}_i$ (see Appendix A); then these numbers were fed
to a Monte-Carlo program which simulated single-electron tunneling events in
the circuit using the ``orthodox'' theory.\cite{Av-Likh} The logical input
was a relatively long (typically, 16-bit) quasi-random bit sequence,
repeated periodically many times.

The simulations have shown that for the geometry presented in Fig.\ \ref
{SET-sr}, elliptic polarization of the clock field (rather than the simple
circular polarization implied above) is essential to provide broad margins.
More exactly, the out-of-plane component $A_z$ should be larger than the
in-plane component $A_y$, because of the field screening in $z$ direction
due to the finite size of the conducting islands. For example, for the
circuit with $r/R=0.3$, $d/R=1.5$, $b/R=0.7$, and the intercell period in 
$y$-direction $l/R=3$, the best ratio $A_z/A_y$ is close to 1.5. If this is
done, sufficiently wide margins for the amplitudes of the bias field are
really achievable. For $A_z/A_y=1.5$, the 18-cell shift register with the
parameters specified above operates correctly (at zero temperature and
sufficiently low clock frequency) within the range $3.25\,<\,A_y/(e/4\pi
\epsilon\epsilon _0R^2)\,<\,5.1$, corresponding to a margin of $\pm 22\%$. If 
now $A_y$ is kept constant at $A_y=4.15\,e/4\pi \epsilon\epsilon _0R^2$, the
allowed range for $A_z/(e/4\pi \epsilon\epsilon _0R^2)$ is between $4.6\,$
and $8.55,$ corresponding to a margin of $\pm 30\%$. These margins are much
broader than those for the first generation of wireless single-electron
devices. \cite{Kor-isl}

The margins for parameters which destroy the geometric symmetry of the
parametron cells are much narrower. Numerical simulations show that for a
shift register with the parameters quoted above, the maximum allowable shift
of the edge islands of a cell in the opposite directions (parallel to the
central island shift) is about $\pm 0.010\,R$. This corresponds to the
maximal fluctuation of $\pm 0.4$ degrees of the alignment of the line
connecting the edge islands of the cell and the $x$-axis. (A shift of only
one island leads to the same effect). Possibly the situation may be improved
by using alternative geometries and charge modes.

\section{SET Parametron Logic Gates}

The shift register described in the previous Section is actually a line of
inverters. To have a complete set for the arbitrary logic functions we need
to have other logic gates (e.g., NAND or NOR), and a circuit with a fan-out
more than one (``splitter'').

All these functions can be naturally implemented using the geometry shown in
Fig.\ \ref{OR/AND}. If the clock field rotation causes the signal
propagation from the bottom to the top we get a fan-out-two circuit, because
the dipole moment of cell $F$ (in its ON state) will determine the charge
state of both cell $A$ and cell $B$, set during their OFF$\rightarrow $ON
switching. On the other hand, if the signal propagates from top to bottom,
we get the implementation of a binary logic function 
(either $F=A.\mbox{NOR}.B$ or $F=A.\mbox{NAND}.B$, 
depending on the asymmetry provided by the background charges or dc 
bias field $\langle E_{s}\rangle $ imposed on cell $F$).

In order to verify the operation of this system, we have carried out its
numerical simulations in the symmetric single-exciton mode, using the same
parameters as above ($r/R=0.3$, $d/R=1.5$, $b/R=0.7$, $l/R=3$) for each of
three shift registers, each of them 6 cells long. The $z$-shift between
upper registers was $6R$, while the $y$-spacing between cells $A$ ($B$) and 
$F$ was $2R$, i.e. slightly less than the spacing $l=3R$ between cells in
each register.

The simulations have shown the following margins for the amplitude
components $A_{y}$ and $A_{z}$ of the clock field (at zero temperature and
relatively low speed). For the fan-out-two gate, at $E_{z}/E_{y}=1.5$, we
get correct operation at $3.45\,<\,A_{y}/(e/4\pi \epsilon \epsilon
_{0}R^{2})\,<\,4.2$, corresponding to a margin of $\pm 9\%$. For the
constant $E_{y}=4.15\,e/4\pi \epsilon \epsilon _{0}R^{2}$, the $z$-component
should be within the range $4.6\,<\,A_{z}/(e/4\pi \epsilon \epsilon
_{0}R^{2})\,<\,6.3$ (the margins of $\pm 16\%$). For the NOR gate, with 
$E_{y}=4.1\,e/4\pi \epsilon \epsilon _{0}R^{2}$, $E_{z}=6.15\,e/4\pi 
\epsilon
\epsilon _{0}R^{2}$ (this operating point is within the margins for both the
shift register and fan-out circuits) correct operation is achieved for the
dc bias field $\langle E_{s}\rangle $ equivalent to the background charge
shift in the range $0.034<\,q_{0}/e<0.053$ 
($q_{0}=q_{0,left}=-q_{0,right}$). For the NAND gate with similar 
parameters the range is $0.035\,<\,-q_{0}/e\,<0.069$. 
The difference between the results for the
ranges (which should coincide because of the layout symmetry) is explained
by the finite set of tested input signal sequences which did not have a
symmetry between logical zero and unity. Hence, the true range cannot be
wider than the common part of two ranges [0.035,\thinspace 0.053], and thus
the real margin is slightly below $\pm 20\%$.

\section{Energy dissipation during OFF$\rightarrow$ON switching}

\label{EnDis}Let us calculate the energy dissipation during the switching 
OFF$\rightarrow $ON of one SET Parametron cell in a fixed signal field 
$E_{s}$.
The switching is illustrated in Fig.\ \ref{switching} which shows the
energies of three possible states of the cell as functions of time. For a
cell charged by one electron the letters $l$, $r$, and $m$ correspond to the
location of the extra electron. For the single-exciton case we can use the
same notation, counting one hole or electron as background charge.

Within the framework of the ``orthodox'' theory the switching is described
by the master equation,\cite{Av-Likh} 
\begin{eqnarray}
&&\frac{d}{dt}\sigma _{l(r)}=\sigma _{m}\Gamma _{l(r)}^{+}-\sigma
_{l(r)}\Gamma _{l(r)}^{-}\,,  \nonumber \\
&&\frac{d}{dt}\sigma _{m}=\sigma _{l}\Gamma _{l}^{-}+\sigma _{r}\Gamma
_{r}^{-}-\sigma _{m}(\Gamma _{l}^{+}+\Gamma _{r}^{+})\,,  \nonumber \\
&&\sigma _{m}+\sigma _{l}+\sigma _{r}=1\,,\,\,\,\sigma _{m}(-\infty )=1\,,
\label{swit-maseq}
\end{eqnarray}
where $\sigma _{m}$, $\sigma _{l}$, $\sigma _{r}$ are the probabilities of
finding the system in ``$m$'', ``$l$'', and ``$r$'' states. The tunneling
rates $\Gamma _{l(r)}^{\pm }$ are given by Eq.\ (\ref{Gamma}), in which we
now assume the linear dependence of the energies on time: 
\begin{equation}
W_{r}^{+}(t)=-W_{r}^{-}(t)\equiv W(t)=\alpha t\,  \label{W(t)}
\end{equation}
\begin{equation}
W_{l}^{+}(t)=-W_{l}^{-}(t)=W(t)-\Delta .  \label{Wswit}
\end{equation}
(This assumption is evidently valid only if the energy difference $\Delta $
due to the signal field in the moment of switching is much less than the
maximal energy due to the rotating bias field.) We also assume that the
signal field makes state ``$r$'' energetically more preferable, $\varepsilon
_{l}-\varepsilon _{r}=\Delta >0$.

If the clock speed and temperature are sufficiently low so that the system
switches correctly from ``$m$'' to ``$r$'' (bit errors will be considered in
Section IX), then the energy dissipation can be calculated considering only
states ``$r$'' and ``$m$'' and neglecting state ``$l$''. In this case the
master equation (\ref{swit-maseq}) is simplified with the condition $\sigma
_{l}(t)=0$, 
\begin{equation}
\frac{d}{dt}\sigma _{r}=(1-\sigma _{r})\Gamma ^{+}-\sigma _{r}\Gamma ^{-}\,
\label{red-maseq}
\end{equation}
(here we omit index $r$ in tunneling rates for simplicity), and has an
explicit solution 
\begin{equation}
\sigma _{r}(t)=\int_{-\infty }^{t}\Gamma ^{+}(t^{\prime })\exp \left[
-\int_{t^{\prime }}^{t}\Gamma _{\Sigma }(t^{\prime \prime })dt^{\prime
\prime }\right] dt^{\prime }\,,  \label{sigmar(t)}
\end{equation}
where 
\begin{equation}
\Gamma _{\Sigma }(t)=\Gamma ^{+}+\Gamma ^{-}=\Gamma ^{+}\left[ 1+\exp
(-W(t)/T)\right] \,.  \label{GammaSigma}
\end{equation}

The statistical average of energy loss by the moment $t$ is given by the
expression 
\begin{equation}  \label{dissip}
{\cal E}(t)=\int_{-\infty }^tW(t^{\prime }) {\dot \sigma }_r(t^{\prime
})dt^{\prime }\, .
\end{equation}
Using Eq.\ (\ref{sigmar(t)}) we find, finally: 
\begin{eqnarray}  \label{dis-sol}
{\cal E}(t)= && \int_{-\infty }^tW(t^{\prime })\Gamma _\Sigma (t^{\prime }) 
\left[ \frac 1{1+\exp (-\frac{W(t^{\prime })}T)}
 \right.  \nonumber \\
&&  -\int_{-\infty }^{t^{\prime }}\Gamma _\Sigma (t^{\prime \prime }) 
\frac 1{1+\exp (-\frac{ W(t^{\prime \prime })}T)}
	\nonumber \\
&& \left. \times \exp [-\int_{t^{\prime
\prime }}^{t^{\prime }}\Gamma _\Sigma (t^{\prime \prime \prime })dt^{\prime
\prime \prime }]\,dt^{\prime \prime }\right] dt^{\prime }\, .
\end{eqnarray}

The solid line in Fig.\ \ref{diss-c-c} shows the total energy loss ${\cal E=}
$ ${\cal E}(\infty )$ as a function of the dimensionless switching speed 
\begin{equation}
\beta \equiv \alpha e^{2}/GT^{2}.  \label{beta}
\end{equation}
In the low-speed limit, $\beta \ll 1$, the switching process is almost
adiabatic. It consists of numerous tunneling events (back and forth between
``$m$'' and ``$r$'' states) occurring within the energy interval on the
order of the thermal energy $T$ around the point $W(t)=0$. (Figure \ref
{switching} shows a particular Monte-Carlo realization for $\beta =0.1$.) It
is easy to check that in the purely adiabatic case when $\sigma _{r}$ is
determined by thermodynamics only, 
\begin{equation}
\sigma _{r}(t)=\sigma _{r}^{ad}(t)\equiv \frac{\Gamma ^{+}}
{\Gamma _{\Sigma }}=\frac{1}{1+\exp (-W(t)/T)}\,,  \label{sig-ad}
\end{equation}
the total dissipation is zero, because in Eq.\ (\ref{dissip}) the symmetric
function of time ${\dot{\sigma}}_{r}(t)$ is multiplied by the antisymmetric
function $W(t)$.

According to Eq.\ (\ref{red-maseq}), the correction to the adiabatic limit, 
$\Delta \sigma _{r}\equiv \sigma _{r}-\sigma _{r}^{ad}$ satisfies the
equation $\Delta \sigma _{r}=-{\dot{\sigma}}_{r}/\Gamma _{\Sigma }$; hence,
as the first approximation in $\beta $ we can use 
\begin{equation}
\Delta \sigma _{r}=-{\dot{\sigma}}_{r}^{ad}/\Gamma _{\Sigma }.
\label{Deltasigma}
\end{equation}
After the integration by parts of Eq.\ (\ref{dissip}) the total dissipation
is given by 
\begin{equation}
{\cal E}=\int_{-\infty }^{\infty }W(t)\,\frac{d\Delta \sigma _{r}}{dt} 
\,dt=-\int_{-\infty }^{\infty }\frac{dW(t)}{dt}\,\Delta \sigma _{r}\,dt\,.
\label{netdis}
\end{equation}
Combining Eqs.\ (\ref{Deltasigma}), (\ref{netdis}), (\ref{W(t)}), (\ref
{GammaSigma}), and (\ref{Gamma}), we obtain the low-speed approximation for
the total dissipation 
\begin{equation}
{\cal E}=\kappa \beta T=\kappa \frac{\alpha e^{2}}{GT}\,,  \label{lowspeed}
\end{equation}
\begin{equation}
\kappa =\int_{-\infty }^{\infty }\frac{e^{-x}(1-e^{-x})}{x(1+e^{-x})^{3}} 
\,dx\,\simeq 0.426\,.  \label{kappa}
\end{equation}
In this quasi-reversible regime ${\cal E}\ll T$; notice also that in this
limit the total dissipation {\it decreases} when temperature {\it increases}.

In the opposite limit, $\beta \gg 1$, the speed of energy change is so high
that the tunneling occurs only at $W>0$ and only once -- see Fig.\ \ref
{switching}, without numerous back and forth processes. (Actually, this
limit is the only one possible at $T=0$.) Then the solution of Eq.\ (\ref
{red-maseq}) is 
\begin{eqnarray}
\sigma _{r}(t) &= &1-\exp \left( -\int_{0}^{t}\Gamma ^{+}(t)dt\right)
	\nonumber \\
& = & 1- \exp
\left( -\frac{\alpha Gt^{2}}{2e^{2}}\right) ,\,\,t>0,
\end{eqnarray}
and the average total dissipation 
\begin{equation}
{\cal E}=(\pi \beta /2)^{1/2}T=(\pi e^{2}\alpha /2G)^{1/2}.
\label{highspeed}
\end{equation}
Hence, in the case $\beta \gg 1$ the average energy dissipation is much
larger than (and independent of) the thermal energy, indicating the
thermodynamic irreversibility of the switching process in this limit. In
Fig.\ \ref{diss-c-c} the low-speed and high-speed approximations (Eqs.\ (\ref
{lowspeed}) and (\ref{highspeed})) are represented by dashed lines, while
the solid line shows the result of the exact calculation using Eq.\ (\ref
{red-maseq}), which gives a natural crossover between these two limits.

Equation\ (\ref{Gamma}) is valid only for the case of the continuous
spectrum of electron energies in the conducting islands. It is easy to
consider a different case when the electron energy in the middle island is
strongly quantized so that only one level is involved in tunneling, while
the energy spectrum of the edge island is still continuous. (This situation
is possible when the middle island is much smaller). Then the ``orthodox''
theory should be somewhat modified, \cite{Av-Kor,Av-Kor-Likh,Beenakker}
however, for our purpose the only important change is that Eq.\ 
(\ref{Gamma}) should be replaced with 
\begin{eqnarray}
&& \Gamma (W)=\frac{\Gamma _{0}}{1+\exp (-W/T)},\,\,
	\nonumber \\
&& \Gamma (W)+\Gamma
(-W)=\Gamma _{0}=\mbox{const},  \label{Gamma-discr}
\end{eqnarray}
where $\Gamma _{0}$ is a constant which characterizes the tunnel barrier
transparency. Equation\ (\ref{dis-sol}) yields that the average total 
energy loss during one switching is given by a simple formula 
\begin{equation}
{\cal E}=\alpha /\Gamma _{0}  \label{diss-discr}
\end{equation}
for arbitrary switching speed (see Fig.\ \ref{fig-discr}). It is evident
that quasi-reversibility (${\cal E}\ll T$) is possible in this case as well,
if the switching speed is low enough: $\alpha \ll T/\Gamma _{0}$.

In both cases considered above, electron energy relaxation in islands with
continuous spectrum (implied by Eqs. (\ref{Gamma}) and (\ref{Gamma-discr}))
is the source of dissipation in the system. In the important case when
electron energy is quantized in {\it all} islands this may not be true, so
that this limit requires a completely different treatment. Our hope is to
complete an analysis of this important case in the near future.

To conclude this section, notice that our model allows not only the total
dissipation ${\cal E}$ to be calculated, but also the time dynamics of the
heat transfer between the system and the environment (``heat bath'') during
the switching process to be followed. Dashed lines in Figs.\ \ref{diss-c-c}
and \ref{fig-discr} show the results of such calculation using Eqs. (\ref
{Gamma}) and (\ref{Gamma-discr}), respectively. During the first part of the
switching process (when $W(t)\leq 0$) the energy ${\cal E}_{1}\equiv -
{\cal E}(0)$ is (on the average) {\it borrowed} from the heat bath which, 
hence, is 
{\it cooled}. During the second part of the process ($W(t)\geq 0$) the
average energy ${\cal E}_{2}\equiv {\cal E}-{\cal E}(0)>{\cal E}_{1}$ is 
{\it returned} back to the heat bath. Notice that in the adiabatic limit 
${\cal E}_{1}={\cal E}_{2}=T\ln 2$ independently of the exact model used 
for $\Gamma (W)$. This result follows from Eqs.\ (\ref{dissip}) and 
(\ref{sig-ad}) after the integration by parts: 
\begin{equation}
{\cal E}_{1}^{ad}={\cal E}_{2}^{ad}=\int_{-\infty }^{0}\frac{1}
{1+\exp (-W/T) }\,{\dot{W}}\,dt=T\ln 2.  \label{e1ad}
\end{equation}
Notice that this is valid for any gradual function $W(t)$, if only $W(0)=0$.

The generality of this result is due to the direct relation ${\cal E}
(t)=-T\Delta S_{inf}(t)$ between the energy and the informational
(``Shannon'') entropy $S_{inf}$ of the degree of freedom used to code
information (for a more detailed discussion, see Appendix B). In fact, in
the instant when $W=0$, the system may be in either of two states ($\sigma
_{m}=\sigma _{r}=1/2)$, hence $\Delta S_{inf}=\ln 2$ has been acquired in
comparison with the definite initial state 
($\sigma _{m}=1$, $\sigma _{r}=0)$. By the end of the switching 
the informational entropy is restored to the
initial value since the state is definite again ($\sigma _{m}=0$, $\sigma
_{r}=1$). Finite switching speed decreases ${\cal E}_{1}$ and increases 
${\cal E}_{2}$ (see the dotted lines in Figs.\ \ref{diss-c-c} and \ref
{fig-discr}) leading to a positive total dissipation ${\cal E}={\cal E}_{2}- 
{\cal E}_{1}$.

\section{Reversible computation}

The general thermodynamic arguments lead to the conclusion that erasure of
information necessarily requires an energy dissipation of at least $T\ln 2$
per bit (see Refs.\ \cite{infentropy,Landauer,Bennett} and Appendix B).
During the switching OFF$\rightarrow $ON of a cell in any SET-Parametron
circuit, the amount of information is not changed, thus allowing arbitrary
small energy dissipation in the slow-switching limit. However, for switching
ON$\rightarrow $OFF the lower bound on dissipation is determined by logical
reversibility of a particular circuit.

The SET Parametron shift register is obviously a logically reversible
circuit, because during cell switching to OFF the information is preserved
by the next cell. It is easy to check (see Fig.\ \ref{SET-sr}) that the sign
of the energy difference $\Delta =\varepsilon _{l}-\varepsilon _{r}$ between
two digital states does not change during the ON phase because of the back
influence from the next cell. So the cell stays in the lower-energy state
and the analysis of the energy dissipation during switching ON$\rightarrow 
$OFF is equivalent to that of the switching OFF$\rightarrow $ON (see Fig.\ 
\ref{ONOFF}). A similar small-dissipation case is realized in the SET
Parametron fan-out circuit, because during the ON$\rightarrow $OFF switching
of the last cell of the input line (cell $F$ in Fig.\ \ref{OR/AND}) the
proper sign of $\Delta $ is maintained by both first cells of the output
lines.

The situation is different for the NAND/NOR gate because any gate consisting
of two inputs and one output is logically irreversible and, hence, has a
lower bound for dissipation. \cite{infentropy,Landauer,Bennett} For two
uncorrelated input streams of bits with equal probabilities of unity and
zero, the informational entropy before the logic operation is 
$S_{inf}^{before}=-\ln (1/4)$ while after the completion of the logic
operation and erasure of the input information it becomes 
$S_{inf}^{after}=-\ln (1/2)$. (Notice that the informational entropy 
{\it decreases}. This seems to contradict the apparent partial loss of
information at computing. Actually the {\it uncertainty} is partially lost,
not the information, because in the Shannon formalism we should treat the
input data as unknown, and the data bit count is decreased by the gate.) The
entropy difference determines the lower bound for the average dissipation 
${\cal E}$ per logic operation ${\cal E}\geq
T(S_{inf}^{before}-S_{inf}^{after})=T\ln 2$.

Actually, in the SET Parametron realization shown in Fig.\ \ref{OR/AND} the
average energy dissipation is much larger than this lower bound. If two
input bits are different, the energy difference $\Delta $ changes its sign
during ON state of either cell A or B. In this case the energy dissipation
during switching to OFF is comparable to $|\Delta |$ (see Fig.\ \ref{ONOFF})
which should be much larger than the thermal energy because the condition 
$|\Delta |\gg T$ is necessary to ensure small error probability (see next
Section).

To realize {\it reversible} NOR and NAND operations (which would provide
small dissipation) using SET Parametron cells, gates with two inputs and
three outputs can be used (see Fig.\ \ref{revers}). This idea was suggested 
\cite{Likh-PQ2} for the Parametric Quantron logic gates and can be directly
applied to the SET Parametron. The input information is copied to the first
cells of two additional shift register lines. If their coupling to the last
cells of input lines is stronger than input-output coupling, the proper sign
of $\Delta $ is always maintained, and the energy dissipation is arbitrarily
small in the slow switching limit.

\section{Bit error rate}

Kinetic equation (\ref{swit-maseq}) allows the calculation of the rate of
``classical'' digital errors during the SET Parametron cell switching (later
we will briefly discuss also the ``nonclassical'' errors due to cotunneling 
\cite{Av-Odin}). The error probability $P$ is given by $\sigma _{l}
(\infty )$. Let us first assume $T=0$ and calculate the ``dynamic'' 
error which occurs
when the switching speed $\alpha $ is too high, and the system remains in
the initial (symmetric) state up to the moment when tunneling to the upper
energy level becomes possible (see Fig.\ \ref{switching} and Eqs.\ (\ref
{W(t)}) -- (\ref{Wswit})). Since there is no tunneling back and forth at 
$T=0 $, the error probability can be found simply as the time integral of 
the rate of erroneous tunneling $\Gamma _{l}^{+}$ multiplied by the 
probability $\sigma _{m}$ that no tunneling has yet occurred: 
\begin{equation}
P_{dyn}=\int_{0}^{\infty }\Gamma _{l}^{+}(\varepsilon )\,\sigma
_{m}(\varepsilon )\,\frac{d\varepsilon }{\alpha }\,,  \label{dyn-gen}
\end{equation}
\begin{eqnarray}
\sigma _{m}(\varepsilon ) =&& \exp \left( -\int_{0}^{\Delta }\Gamma
_{r}^{+}(\varepsilon ^{\prime })\frac{d\varepsilon ^{\prime }}{\alpha }
\right)
	\nonumber \\
&& \times  \exp \left( -\int_{0}^{\varepsilon }\left( \Gamma
_{l}^{+}(\varepsilon ^{\prime })+\Gamma _{r}^{+}(\varepsilon ^{\prime
}+\Delta )\right) \frac{d\varepsilon ^{\prime }}{\alpha }\right) .
\label{dyn-gen2}
\end{eqnarray}

For the ``orthodox'' model of the tunneling rate given by Eq.\ 
(\ref{Gamma}), 
\begin{equation}
P_{dyn}=K\gamma \exp (-\frac{1}{2\gamma })\,,  \label{dyn}
\label{Pdyn}\end{equation}
\begin{eqnarray}
K &=&\frac{1}{2\gamma }-\frac{\sqrt{\pi }}{4\gamma ^{3/2}}\exp (\frac{1}
{4\gamma })[1-\mbox{Erf}(\frac{1}{2\sqrt{\gamma }})]  \nonumber \\
&=&1+\sum_{n=1}^{\infty }\frac{(2n+1)!}{n!}\,(-\gamma )^{n},  \label{dyn-K}
\end{eqnarray}
where $\gamma \equiv \alpha e^{2}/G\Delta ^{2}$. In order to keep 
$P_{dyn}\ll 1$, $\gamma $ should be much less than 1, then $K\approx 1$. 
Equation (\ref{Pdyn}) shows that the dynamic error decreases 
exponentially with the
decrease of the switching speed and even faster than exponentially with the
increase of $\Delta $ (factor $\Delta ^{2}$ in the exponent).

For sufficiently small $\alpha $ and/or large $\Delta $ the main
contribution to the error probability will be due to the thermally activated
processes which populate the symmetrical state ``$m$'' during the passage of
energy $\varepsilon _{m}$ across $\varepsilon _{l}$. The probability of this
``thermal'' error is given by the simple formula 
\begin{equation}
P_{therm}=\exp (-\Delta /T)\,  \label{therm-err}
\end{equation}
and it prevails over $P_{dyn}$ when $T\gg \alpha e^{2}/G\Delta $. In the
case when both errors are of the same order, the result can be found by the
numerical solution of Eq.\ (\ref{swit-maseq}). The total error probability
can be estimated as the maximum of the two analytical results presented
above: $P\simeq \max (P_{dyn},\,P_{therm})$.

If instead of the ``orthodox'' model we assume that only one discrete energy
level of the middle island participates in tunneling, then using Eq.\ (\ref
{Gamma-discr}) for the tunneling rate we obtain from Eqs.\ (\ref{dyn-gen})
and (\ref{dyn-gen2}) the following probability of the dynamic error ($T=0$): 
\begin{equation}
P_{dyn}=\frac{1}{2}\exp (-\frac{\Delta }{\alpha }\Gamma _{0})\,.
\label{dyn-discr}
\end{equation}
The thermal error is still given by Eq.\ (\ref{therm-err}), and it prevails
over the dynamic error if $T\gg \alpha /\Gamma _{0}$.

One more possible source of errors is the higher-order quantum process of
cotunneling \cite{Av-Odin} when two or more electrons tunnel simultaneously
through different junctions. For illustration, the lowest energy diagram in
Fig.\ \ref{three-isl}b shows the situation when the charge state with higher
energy is occupied, and the digital information in the cell is preserved by
the energy barrier (higher energy of the symmetric state) due to the bias
field. According to the orthodox theory, single-electron tunneling in this
case is impossible (at sufficiently low temperature). However, the
second-order cotunneling, i.e. simultaneous tunneling of two electrons
(through both junction) brings the system into the lower energy state and,
hence, is energetically allowed. This process changes the sign of the cell
dipole moment and can lead to the digital error.

This type of error can occur, for example, in the 3-phase shift register
during the phase when the bit is stored by only one cell, and the long-range
interaction with cells carrying other bits (nearest cells are OFF) provides
uncontrolled sign of the energy difference between ``$l$'' and ``$r$''
states. The erroneous bit will then propagate along the shift register.

Though a detailed analysis of cotunneling was not a goal of this work, we
should notice that several means are readily available to reduce the
resulting error probability. First, because the rate of $m$-electron
cotunneling scales as $(GR_{Q})^{m}$ (where $R_{Q}$ $\approx 6.5$ $k\Omega $
is the quantum unit of resistance) while the single-electron rate is
proportional to the first power of $G$, the decrease of $G$ decreases the
relative importance of the cotunneling processes. Another, more powerful
method is to increase the smallest order $m$ of possible cotunneling
processes. This can be done, for example, by increasing the number of
islands per cell. If the cell consists of 5 islands and the ON state
corresponds to an electron on one of the outer islands, then at least the
4-th order cotunneling ($m=4$) is necessary to switch between logical zero
and unity; a further increase of the number of islands per cell makes the
cotunneling rate negligible even for relatively large tunnel conductance.
Finally, it is possible to increase the minimal cotunneling order while
still using three-island cells, by increasing the number of cells which
store the same bit. This can be done, for example, by the increase of the
number of phases in the operation of the shift register. In the realization
shown in Fig.\ \ref{SET-sr} this may easily be achieved by a decrease of the
angle $\theta $ between the planes of neighboring cells. If the bit is
stored by $k$ neighboring cells then the error can occur only if all these 
$k $ cells will simultaneously change their polarizations and if the final
state has a lower energy (any ``partial'' switching would cost at least one
cell-cell interaction energy and, hence, is thermodynamically forbidden at
least for not too large $k$). So, the lowest order of erroneous cotunneling
is $2k$, and the linear increase in ``hardware'' allows the exponential
reduction of the error probability. This method is applicable to any logic
gates which use shift registers as their input and output lines.

\section{Discussion}

In this paper we have presented a functionally complete logic family based
on the SET Parametron cells, including inverter, fan-out-two, NAND and NOR
gates. All the circuits may operate correctly in a common range for the bias
field amplitude. The important advantages of the SET Parametron logic are
wireless operation and extremely low power dissipation possible in the
quasi-reversible mode of operation, with the energy loss much less then 
$k_{B}T\ln 2$ per bit.

While the realization of room temperature operation of the SET parametron
logic requires sub-one-nanometer fabrication technology which is hardly
available at the present time, the operation of simple SET Parametron
circuits (with clock wires) can be readily demonstrated experimentally at
lower temperatures. An obvious choice is the aluminum shadow-mask
evaporation technology, widely accepted for singe-electron device
fabrication (see, e.g., the collection \cite{SCT}) . Figure \ref{alumin}
shows the sketch of a possible layout of the two-cell SET Parametron shift
register-inverter. (For the sake of simplicity, the artifact islands typical
for this technology are not shown). Each cell consists of three islands. The
capacitive gates are used for the application of rf ``clock'' fields which
cause the switching processes, and simultaneously for the application of dc
fields to compensate random background charges. Input gates A and B (for the
initial demonstration, one gate is sufficient) determine the polarization of
the first cell during its OFF$\rightarrow $ON switching. This polarization
in turn determines the polarization of the next cell during its 
OFF$\rightarrow $ON switching. The final polarization of the latter cell 
is sensed by a capacitively coupled single-electron transistor. (If a
multilayer fabrication is available, this layout may be further improved to
provide wider parameter margins: an overlap (without tunneling) of the edge
islands of two cells would increase their capacitive coupling, while a
similar overlap of the bias gates with the islands would allow smaller
cross-talk.) The operation temperature of such a SET-Parametron circuit with
100-nm-scale tunnel junctions will be in sub-1-K range.

Coming back to the wireless realization, let us estimate the parameters of a
possible implementation of the device using the conducting (e.g., metallic)
clusters as islands. For a cluster diameter $2R$ about 5 nm (which is at the
limit of present-day direct e-beam writing techniques) the charging energy 
$E_{c}\sim e^{2}/8\pi \epsilon \epsilon _{0}R$ is about $0.15eV$ (where 
$\epsilon \sim 2$ is taken as a typical dielectric constant for the organic
materials which are the natural candidates for the cluster environment). For
the layouts considered in this paper the typical energy difference $\Delta $
between ``$l$'' and ``$r$'' states is about $0.2E_{c}$ leading to $\Delta
\sim 0.03$ eV. Requiring the probability of the thermal error to be less
than $10^{-10}$ per switching, Eq.\ (\ref{therm-err}) yields the maximal
operation temperature of about 15 K. Assuming the same value for the dynamic
error (then $\gamma \simeq 0.025$ -- see Eq.\ (\ref{dyn})) and taking into
account the particular geometry (Fig.\ \ref{SET-sr}), we obtain the maximum
clock frequency $\nu _{max}\sim 5\times 10^{-4}G\Delta /e^{2}$ that
corresponds to about $10^{9}$ Hz for our parameters and $G\sim $(10$^{5}$
Ohm)$^{-1}$ (higher $G$ would make the cotunneling a problem). In this case
the power dissipation (see Eq.\ (\ref{highspeed})) is about $4(\pi \gamma
/2)^{1/2}\nu \Delta \sim \nu \Delta \sim 5\times 10^{-12}$ W per cell. This
extremely low dissipation would make possible an integration level up to 
10$^{11}$ cells per cm$^{2}$ (i.e.\ 10$^{3}$ nm$^{2}$ per cell), limited by 
the cell size, since at 15 K the heat flux about 1W/cm$^{2}$ can be easily
managed without the circuit overheating. To achieve the quasi-reversible
mode of operation the frequency should be even lower: $\nu \ll 5\times
10^{7} $ Hz for our set of parameters at $T=15$ K. Accepting, for example, 
$\nu =10^{6}$ Hz we get a power dissipation of only about $5\times 10^{-18}$ 
W per cell. This figure indicates that as far as power dissipation is
concerned, three-dimensional integration of SET Parametron circuits is quite
feasible. Despite the not very spectacular clock frequency in this regime,
the total computing performance of a 3D system can be very large.

The largest problem with SET Parametron circuits (besides their fabrication)
is that their operation requires well-defined background charges. The
allowed fluctuations are only on the order of $0.01e$ (this number
corresponds to the maximum tolerated angle of the cell tilt -- see Section
V). This is a common problem for any single-electron logic (see for example
Refs.\ \cite{Kor-rev,Kor-trans,Chen}). However, if/when a molecular technology
becomes available for the implementation of single-electron and other
nanoscale devices, SET Parametron cells may be reproducible on the molecular
level, and well defined background charges seem to be achievable in
principle.

From the estimates above we see that in order to increase the operation
temperature of SET Parametron circuits up to 300 K or even up to 77 K,
sub--nanometer-scale islands are necessary. In this case the energy level
discreteness \cite{Av-Kor,Av-Kor-Likh,Beenakker} may result in radically new
features of the SET Parametron operation, and should be taken into account
at its quantitative analysis. Such analysis is outside the scope of the
present paper. We would just like to notice in passing that the operation of
SET Parametron circuits in that mode would be rather close to the recently
proposed new version \cite{Lent-Tou} of the so-called ''Quantum Cellular
Automata''.\cite{GSC} This new version, put forward under the keywords of
``adiabatic switching'' and ``pipelining'', may remove the principal
difficulties \cite{Kor-isl} of the initial suggestion. \ We are not,
however, aware of any detailed analysis of this new family of logic devices.

\acknowledgements
We would like to thank D. V. Averin and T. Usuki for useful discussions. The
work was partially supported by AFOSR Grant No. F49620-95-1-0044, Russian
RFBR Grant No. 97-02-16332, and Russian Program on 
Nanoelectronics Grant No. 039-03-232/78/4-3.

\appendix
\section{}

In this Appendix we discuss the method of multiple electrostatic images used
to calculate the inverse capacitance matrix $C^{-1}$ and vectors $D$ (island
dipole moments) which describe the influence of the external electric field
for the arbitrary system of conducting spheres. This method was also used
for numerical calculations in another work. \cite{Kor-isl}

It is well known that the electrostatic field of a point-like charge $q$
located at a distance $d$ from the center of an uncharged sphere having the
radius $r$, may be treated as the net field of the charge $q$ and a pair of
fictitious charges located inside the sphere, in free space. The image
charge $-qr/d$ is located at the distance $r^{2}/d$ from the sphere center
(in the same direction as the charge $q$), and the compensating charge $qr/d$
is located at the sphere center. The total field of these three charges
makes the sphere surface equipotential with $\phi =q/4\pi \epsilon _{0}d$.

Let us use this method to calculate the inverse capacitance matrix of two
spheres with radii $r$ and $R,$ with distance $d$ between the centers. For
that, we need to calculate the electrostatic potential when a charge $q$ is
placed on the sphere $R$. First, let us create the pair $\mp qr/d$ inside
the sphere $r$ (as if we had a point-like charge $q$). Three charges provide
the equipotentiality of the sphere $r$, but not sphere $R$. So, we need to
create two more charge pairs inside sphere $R$ to restore its
equipotentiality. One pair (image of the image charge) will have charges 
$\pm qd/r\times R/(d-r^{2}/d)$ and the other pair (image of the compensating
charge) will have charges $\mp qd/r\times R/d$ (notice that one charge from
each pair is positioned at the center of sphere $R$). Now the
equipotentiality of the sphere $r$ is lost again and we need to create four
new pairs inside it (actually, three pairs because the position of two
charges at the center of $R$ coincides). We can continue this procedure
until the charges of new pairs become sufficiently small, and the sphere
potentials calculated at each iteration converge with the required accuracy.

The same method is trivially generalized to an arbitrary number of spheres:
we need to make images of all charges in all spheres. The method is very
simple mathematically but cannot be applied in a straightforward way if the
number $N$ of spheres is large, because the number of required image pairs
scales as $N^{n}$ where $n$ is the number of iterations. If $N\sim 100$ and 
$n\sim 20$ (that is a typical number of necessary iterations if the spheres
are close to each other, and we require a good accuracy), then the computer
memory will obviously be insufficient.

To solve this problem we use three ways (levels) of storage of information
about the pairs, depending on their magnitude (the pair magnitude can be
characterized by two numbers: the distance between the charges and the
dipole moment of the pair). For ``large'' pairs we store the whole
information (position and the charge magnitude). When the image of a
sufficiently ``small'' pair is calculated, we store only the location and
the magnitude (dipole moment) of the dipole consisting of two image charges,
and the total charge at the sphere center. The lowest level of the
information representation is to consider the dipole being located at the
center of the sphere, so we can sum up all the dipole moments.

This modification of the algorithm allows us to calculate the capacitance
matrix for a few hundred spheres with a typical distance between the spheres
as small as one tenth of the sphere radius using a modest personal computer.
(More closely located spheres require a larger number of image charges.)

The calculation of the $D$ vectors is analogous to the calculation of the
inverse capacitance matrix. For example, to determine the influence of 
$E_{x} $ field we calculate first the dipole moments (which will have only 
$x$-component) produced by the field in the independent-sphere approximation. 
Then we use the iteration procedure described above to restore the
equipotentiality of all spheres. At this stage image dipoles (with all
spatial components) as well as image charge pairs appear. The island
potentials for the unit external field $E_{x}$ are the components of the $D$
vector corresponding to $x$-axis (we need three $D$ vectors for three
dimensions).

	\section{}

In this Appendix we discuss the heat transfer during the adiabatic
transition (switching) between an arbitrary number of charge states. This
analysis can be used for the fan-out circuits and for SET Parametron cells
consisting of more than three islands. The formalism was developed long ago 
\cite{infentropy,Landauer,Bennett} and constitutes the basis of the
reversible computation analysis giving the lower bound for the energy
dissipation for irreversible logical gates.

The infinitesimal heat transfer to the thermal bath can be written as 
\begin{equation}
dQ=\sum_{i,j}(\varepsilon _{i}-\varepsilon _{j})\,\sigma _{i}\,\Gamma
_{i\rightarrow j}\,dt=-\sum_{i}\varepsilon _{i}\,d\sigma _{i}\,,  \label{dQ}
\end{equation}
where $\varepsilon _{i}$ is the (free) energy of the state $i$, $\sigma _{i}$
is its probability, $d\sigma _{i}$ is the probability change during the
interval $dt$, and $\Gamma _{i\rightarrow j}$ is the rate of transition from
state $i$ to state $j$ (i.e. the corresponding tunneling rate). In the
thermodynamically reversible adiabatic limit the probabilities satisfy the
equilibrium distribution 
\begin{equation}
\sigma _{i}=\sigma _{i,ad}=Z^{-1}\exp (-\varepsilon
_{i}/T),\,\,Z=\sum_{i}\exp (-\varepsilon _{i}/T).  \label{equilib}
\end{equation}
Using the definition of Shannon's informational entropy \cite{infentropy} 
\begin{equation}
S_{inf}=-\sum_{i}\sigma _{i}\,\ln \sigma _{i}\,,  \label{Sinf}
\end{equation}
and using the evident equation $\sum_{i}d\sigma _{i}=0$ we get 
\begin{equation}
dS_{inf}=-\sum_{i}d\sigma _{i}(1+\ln \sigma _{i})=T^{-1}\sum_{i}\varepsilon
_{i}\,d\sigma _{i}.  \label{dS}
\end{equation}
We see that in the adiabatic case 
\begin{equation}
dQ=-T\,dS_{inf}.  \label{dSdQ}
\end{equation}
It is not difficult to prove that in the general (nonadiabatic) case $\delta
Q\geq -T\,dS_{inf}$. Let us introduce the difference 
\begin{equation}
X\equiv \frac{\delta Q}{T}+dS_{inf}=-\sum_{i}d\sigma _{i}\ln \frac{\sigma
_{i}}{\sigma _{i,ad}}\,.  \label{A}
\end{equation}
Using the general master equation 
\begin{equation}
\frac{d\sigma _{i}}{dt}=\sum_{j}\sigma _{j}\Gamma _{j\rightarrow i}-\sigma
_{i}\sum_{j}\Gamma _{i\rightarrow j}  \label{gen-maseq}
\end{equation}
to substitute $d\sigma _{i}$ in Eq.\ (\ref{A}), making the resulting
expression symmetric over indices $i$ and $j$, and using the general
thermodynamic relation for the tunneling rates $\Gamma _{i\rightarrow
j}/\Gamma _{j\rightarrow i}=\exp ((\varepsilon _{i}-\varepsilon _{j})/T)$
(c.f. Eqs.\ (\ref{GammaSigma}) and (\ref{Gamma-discr})) we obtain 
\begin{eqnarray}
 X/dt =&& -\sum_{i,j}\Gamma _{i\rightarrow j}\left[ -\sigma _{i}\ln 
\frac{\sigma _{i}}{\sigma _{i,ad}}
\right.  
	\nonumber \\
&&   +\sigma _{j}\ln \frac{\sigma _{i}}
{\sigma _{i,ad}}\exp (\frac{\varepsilon _{j}-\varepsilon _{i}}{T})
	\nonumber \\
&& \left. 
 -\sigma _{j}\ln \frac{\sigma _{j}}{\sigma _{j,ad}}\exp 
(\frac{\varepsilon _{j}-\varepsilon _{i}}{T})
+\sigma _{i}\ln \frac{\sigma _{j}}{\sigma _{j,ad}} 
\right] .  \label{Adt}
\end{eqnarray}
After the simple transformations and using Eq.\ (\ref{equilib}) we finally
get the expression 
\begin{eqnarray}
X/dt= && -\sum_{i,j}\Gamma _{i\rightarrow j}\ln \left( \frac{\sigma _{i}}
{\sigma_{j}\exp (\frac{\varepsilon _{j}-\varepsilon _{i}}{T})}\right) \,
	\nonumber \\
&& \times \left(
\sigma _{j}\exp (\frac{\varepsilon _{j}-\varepsilon _{i}}{T})-\sigma
_{i}\right)  \label{Afin}
\end{eqnarray}
which is obviously positive or zero. This proves the general inequality 
\begin{equation}
\delta Q\geq -T\,dS_{inf}\,.  \label{dQgen}
\end{equation}
Notice that if we introduce also the ``usual'' entropy $\tilde{S}$ so that 
$\delta Q=T\,dS$, then the total entropy $S_{t}=S_{inf}+\tilde{S}$ is
constant in the adiabatic case. In the general case, the total entropy is a
non-decreasing function of time, $dS_{t}/dt\geq 0$, while the decrease of 
$\tilde{S}$ is not forbidden.

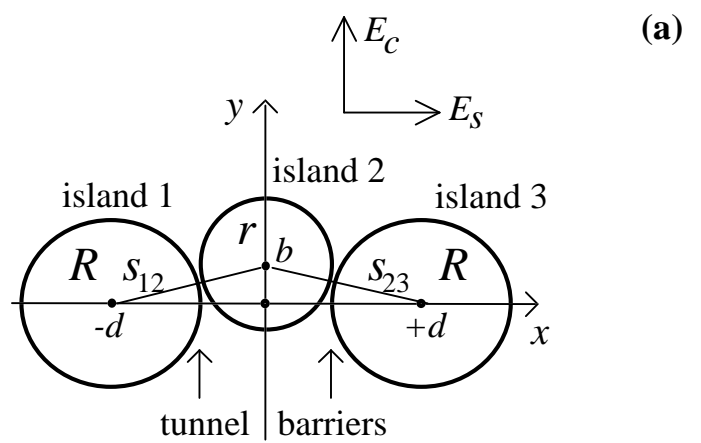
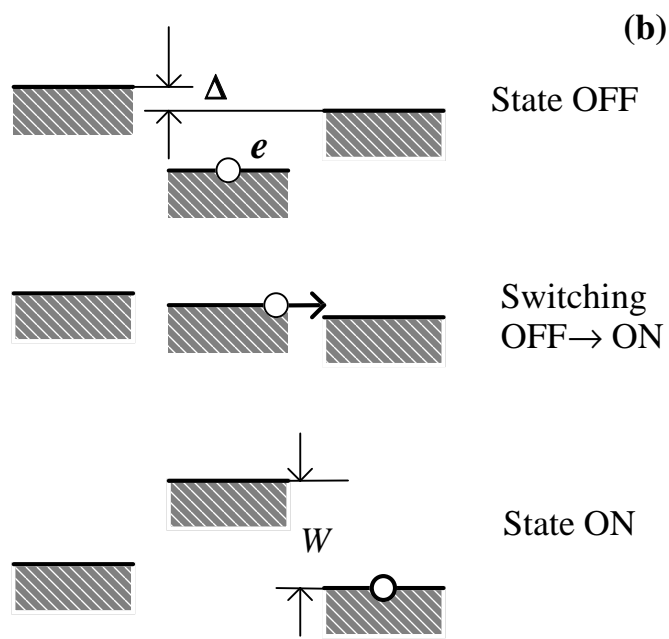
\begin{figure}
\caption{ SET Parametron: (a) the basic cell and (b) its energy diagrams for
three values of the bias field $E_{c}$ (for discussion, see the text).}
\label{three-isl}
\end{figure}

\begin{figure}
\caption{ Phase diagrams of stable charge states of a SET Parametron cell
with $q_{i0}=0$, $R/r=1$, $d/r=3$, $b/r=1$, where $2d$ is the distance
between outer island centers while $b$ is the $y$-axis shift of the middle
island center -- see Fig.\ \protect\ref{three-isl}a. 
(a) The cell is charged with
one extra electron; (b) electrostatically neutral cell. $\Delta E_{c}$ and 
$\Delta E_{s}$ are the diagram periods for the clock and signal electric
fields, respectively. Bistable regions are shaded. Thick line in (a)
illustrates the switching from OFF to ON states, while rectangles in (b)
illustrate two possible operation modes of the cell.}
\label{diag}
\end{figure}

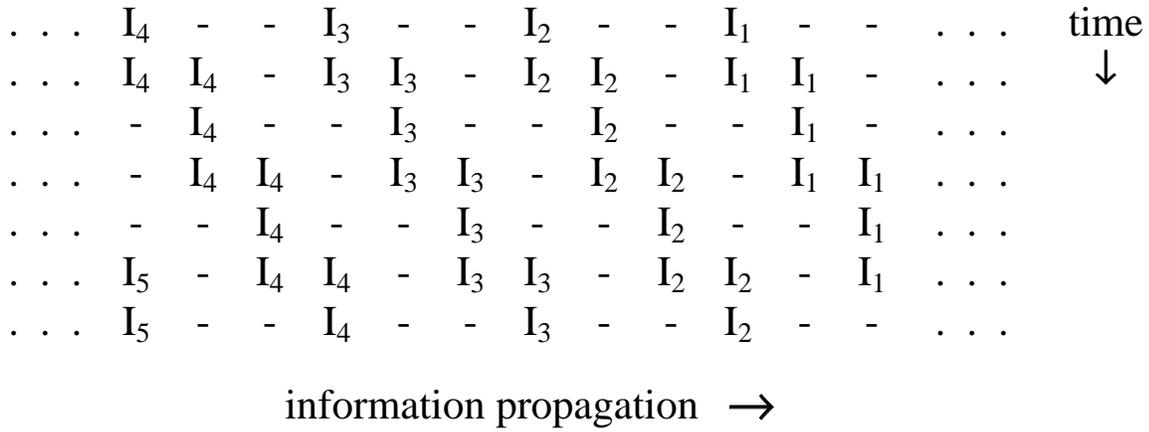
\begin{figure}
\caption{ The general idea of the three-phase shift register using
parametron-type cells. Each line shows the logical state of the shift
register at a particular phase of the clock cycle, dashes indicating the
monostable OFF state and {\bf I} representing bistable ON states.}
\label{SR-idea}
\end{figure}

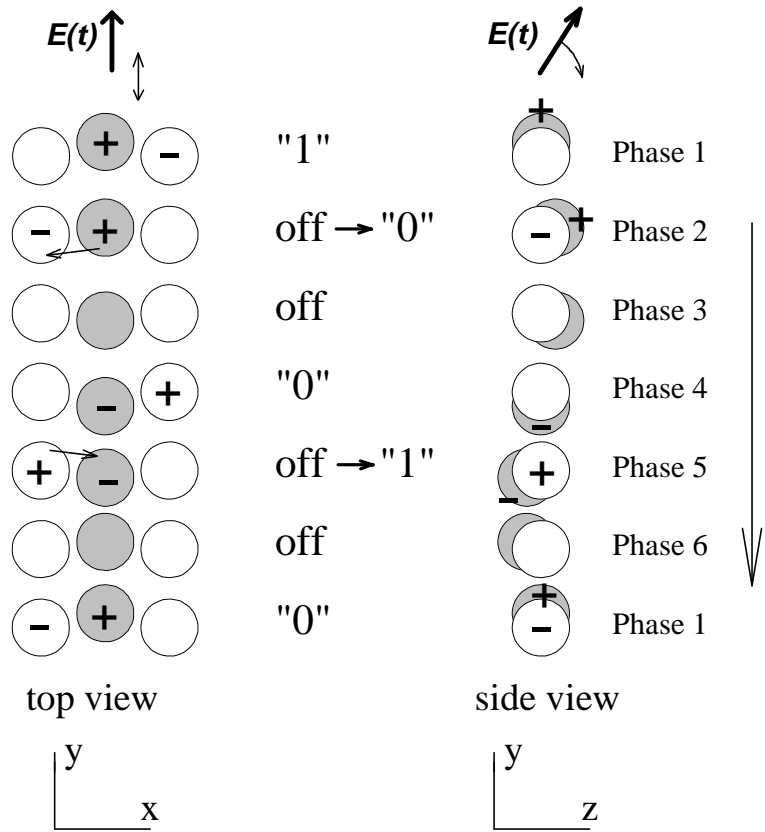
\begin{figure}
\caption{ The top (left) and side (right) views of a shift register based on
the single-exciton parametron cells operating in the symmetric mode. The
clock field vector $E(t)$ rotates in plane $y-z$. Digital bits are coded by
the direction of the cell polarization and are propagated from the top of
the figure to the bottom, over 6 cells during one clock period.}
\label{SET-sr}
\end{figure}

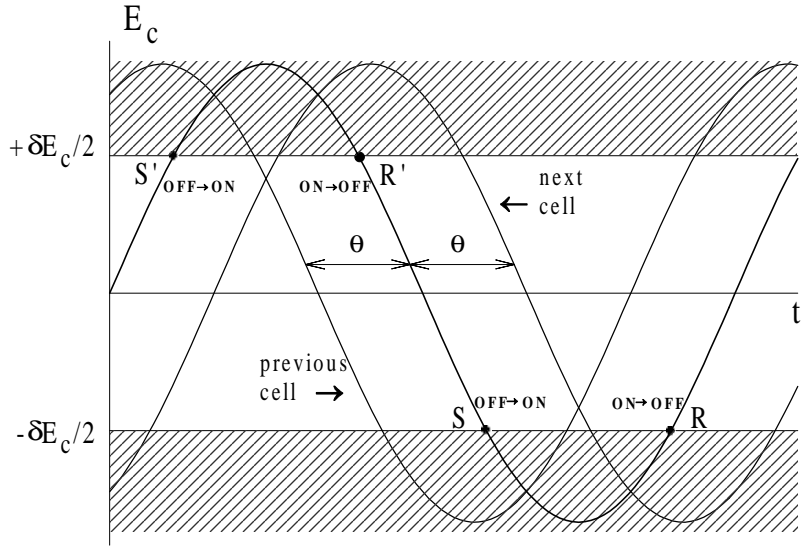
\begin{figure}
\caption{ Time diagram illustrating the operation of the shift register
shown in Fig.\ \protect\ref{SET-sr}. 
Three sine curves show the in-plane component
of the clock field for three neighboring cells. Dots correspond to the
points where the middle curve enters or exits the bistable ON regions
(shaded), so that the middle cell changes its charge state.}
\label{sr-diag}
\end{figure}

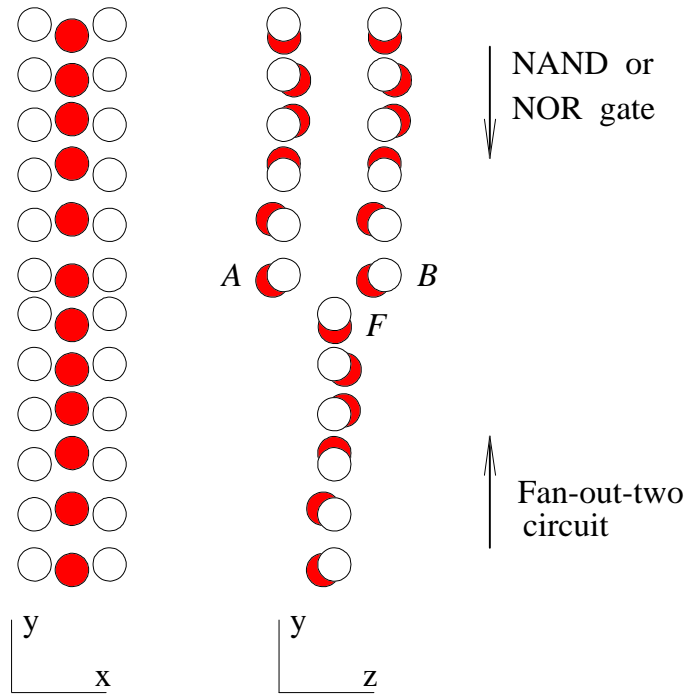
\begin{figure}
\caption{ Circuit which can be used for the fan-out-two splitting of a
signal (if the propagation direction is from the bottom to the top), or as a
logical gate NAND or NOR (for opposite signal propagation direction). The
asymmetry required for NAND and NOR gates can be created by the adjustment
of the background charges on the edge islands of cell $F$, or by application
of local electrostatic field $\langle E_{s}\rangle $.}
\label{OR/AND}
\end{figure}

\begin{figure}
\caption{ Energy diagram of the OFF$\rightarrow $ON switching. The system,
initially in state ``m'', is switched to state ``r'' (switching to ``l''
would give a digital error). Numerous tunneling events (back and forth)
occur in the slow switching case $\protect\beta \ll 1$ (a particular result
of Monte-Carlo simulation for $\protect\beta =0.1$ is shown), while for 
$\protect\beta \gg 1$ there is only one tunneling event. }
\label{switching}
\end{figure}

\begin{figure}
\caption{ Components of the energy exchange between the parametron and the
heat bath as functions of the process speed $\protect\alpha =dW/dt$. Dotted
lines: average energy flow ${\cal E}_{1}$ from the heat bath to the
parametron during the first part of the process (when $W\leq 0$), and the
average flow ${\cal E}_{2}$ from the device back into the heat bath during
its second part (when $W\geq 0$). Solid line: the net energy dissipation 
${\cal E}={\cal E}_{2}-{\cal E}_{1}$. Dashed lines show the low-speed (Eq.\ 
(\protect\ref{lowspeed})) and high-speed (Eq.\ (\protect\ref{highspeed})) 
asymptotes. }
\label{diss-c-c}
\end{figure}

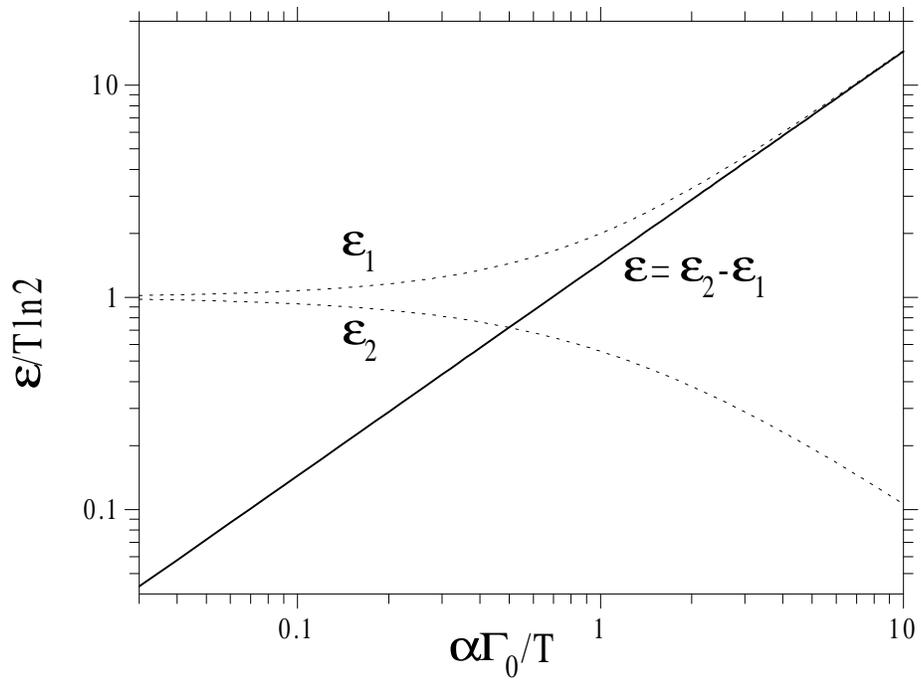
\begin{figure}
\caption{The same as in Fig.\ \protect\ref{diss-c-c}, but for the case of the
discrete energy spectrum of the middle island.}
\label{fig-discr}
\end{figure}

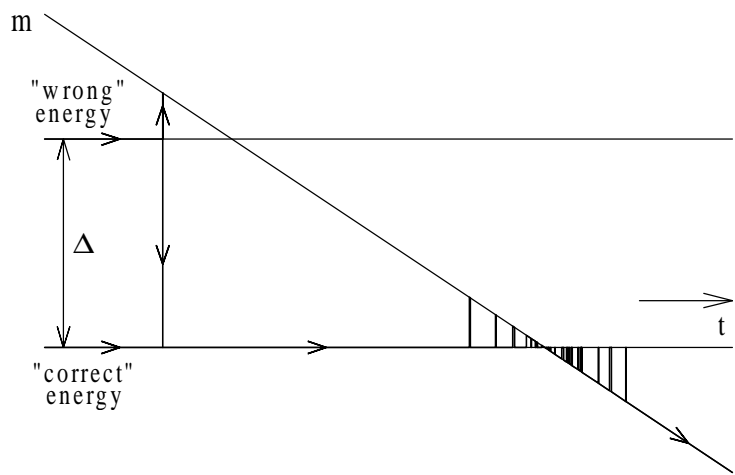
\begin{figure}
\caption{ Energy diagram of the ON$\rightarrow $OFF switching. If the system
initially occupies the state with lower energy, arbitrary small energy
dissipation is possible at slow switching (simulation for $\beta 
=0.1 $ is shown). On the contrary, if the switching starts from the upper
state, the energy loss is finite and of the order of $|\Delta |$. (For this
case, Monte Carlo simulation has given two almost simultaneous tunneling
events shown by arrows).}
\label{ONOFF}
\end{figure}

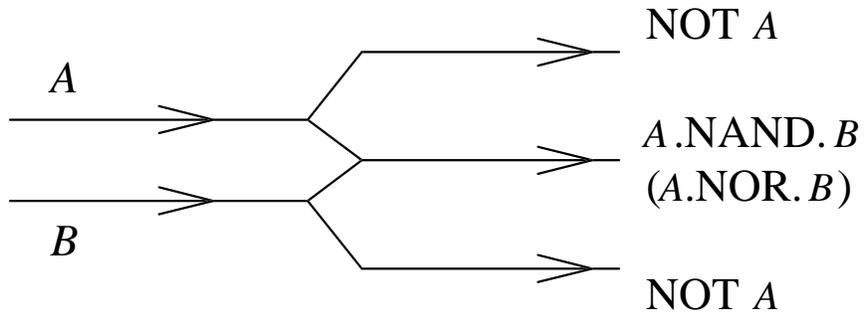
\begin{figure}
\caption{Schematics of logically reversible NAND/NOR gate with two input
shift-register lines and three output lines. The input information is
preserved in two additional output lines.}
\label{revers}
\end{figure}

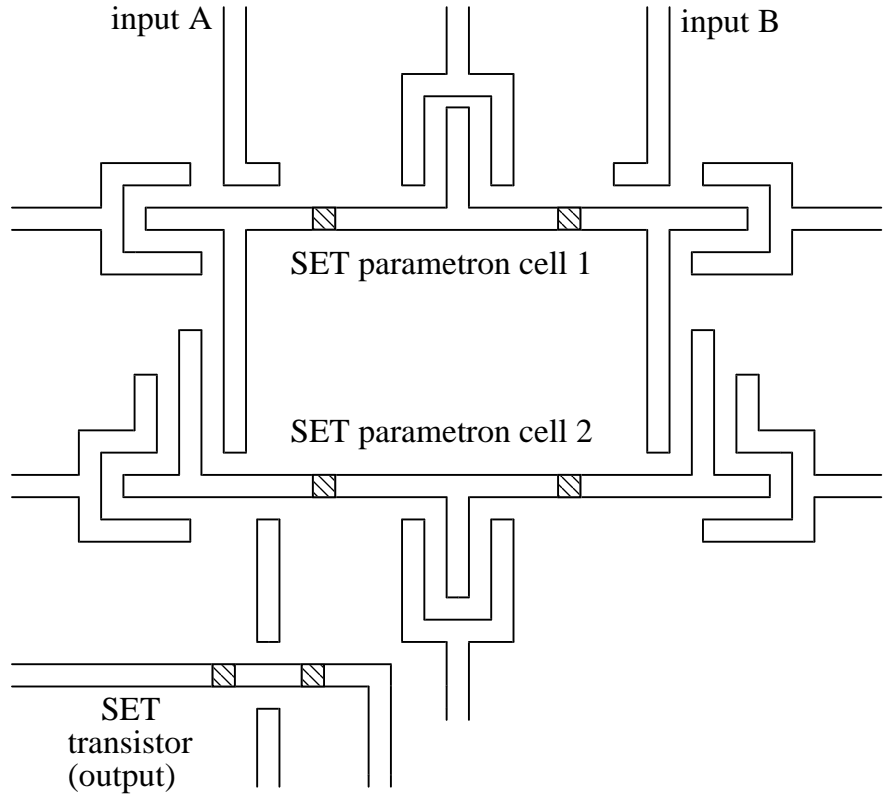
\begin{figure}
\caption{Sketch of a possible layout of the two-cell SET-Parametron shift
register suitable for the standard double-angle shadow evaporation technique
(the artifact islands are not shown).}
\label{alumin}
\end{figure}


\begin{references}
\bibitem{sign}  For the sake of simplicity and following the tradition of
single-electronics, we will assume that the charge of the uncompensated
charge is positive, still referring to it as an electron. This will not
affect any finite result.
\bibitem{higher-states}  In principle, an electron can also be in different
states (of higher energy) inside each island, but these states are not
explicitly participating in the theory of single-electron tunneling, at
least in its simplest ``orthodox'' version. \cite{Av-Likh}
\bibitem{inversion}  Using approximation (\ref{approx}) one could expect
that for sufficiently large ratio $r/R$ it should be possible to satisfy
inequality $2/s-1/r-1/2d>0$ which would lead to $\Delta E_{c}<\delta E_{c}$
(simultaneously $\Delta E_{s}>\delta E_{s}$). This would ``rotate'' the
phase diagram by 90 degrees, reversing the roles of $E_{s}$ and $E_{c}$.
However, this conclusion is an artifact of approximation (\ref{approx})
which is too inaccurate when $s-R-r\alt 
\max (R,r)$. Actually the axis reversal described above is impossible since
the required relation $2C_{12}^{-1}-C_{22}^{-1}-C_{13}^{-1}>0$ contradicts
the general electrostatic property 
$C_{13}^{-1}>C_{12}^{-1}C_{23}^{-1}/C_{22}^{-1}$. The latter inequality can
be derived calculating the potential of the third island due to the positive
charge on the first island with the grounded second island and requiring the
result to be positive.
\bibitem{doubling}  Notice that because of the double switching per period,
the speed of the data propagation is 6 cells per clock cycle, i.e. twice as
large as the number of cells per bit (3). There is one more difference
between the implementation of the shift register (Fig.\ \ref{SET-sr}) and
the general scheme shown in Fig.\ \ref{SR-idea}: the repulsion of similar
electric charges leads to the inversion of the sign of the dipole moment at
each register step. Hence the correspondence between the digital bit and the
electric moment alternates each next cell.
\end{references}
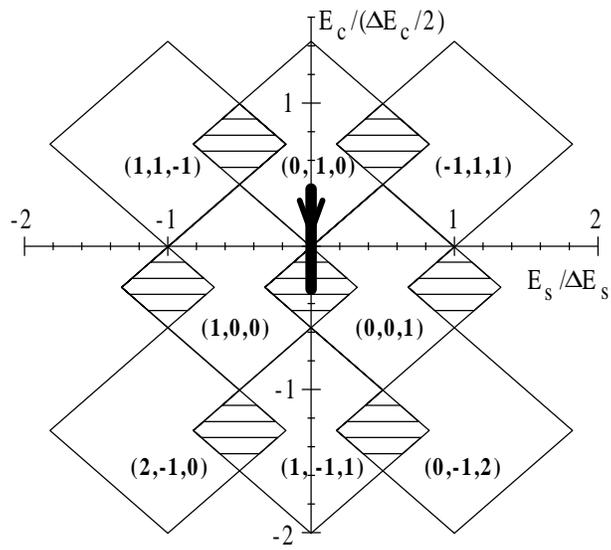
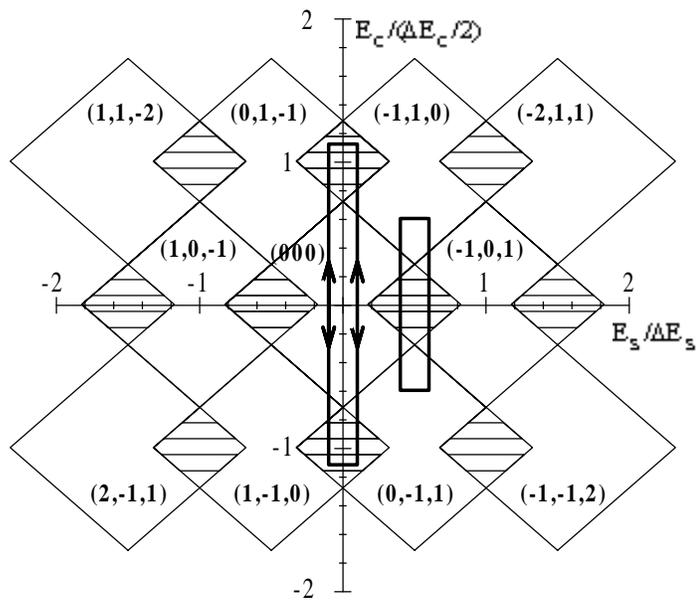
\end{document}